# HIGH-DIMENSIONAL ENCODING COMPUTATIONAL IMAGING

**YONGKANG YAN,[1,2#] ZEQIAN GAN,[1#] LUYING HU,[1] XINRUI XU,[1] RAN KANG,[3] CHENGWEI QIAN,[1] JIANQIANG MEI,[1,2] PAUL BECKETT,[5] WILLIAM SHIEH,[6] RUI YIN,[1,2] XIN HE,[1,2,4*] AND XU LIU[4*]**

[1]School of Information and Electrical Engineering, Hangzhou City University, Hangzhou, 310015, China

[2]School of Information and Software Engineering, East China Jiaotong University, Nanchang, 330013, China

[3]School of Art and Archaeology, Hangzhou City University, Hangzhou, 310015, China

[4]State Key Laboratory of Modern Optical Instrumentation, College of Optical Science and Engineering, Zhejiang University, Hangzhou, 310027, China

[5]STEM College, RMIT University, Melbourne, VIC 3000, Australia

[6]School of Engineering, Westlake University, Hangzhou, 310010, China

[*]liuxu@zju.edu.cn

[*]hexin@hzcu.edu.cn

**Abstract**

High-dimensional imaging technology has demonstrated significant research value across diverse fields, including environmental monitoring, agricultural inspection, and biomedical imaging, through integrating spatial (X*Y), spectral, and polarization detection functionalities. Here, we report a High-Dimensional encoding computational imaging technique, utilizing 4 high-dimensional encoders (HDE1-4) and a high-dimensional neural network (HDNN) to reconstruct 80 high-dimensional images of the target. The system efficiently acquires spectral-polarization information, spanning a wavelength range of 400-800 nm at intervals of 20 nm, obtaining 20 spectral datasets. Each dataset contains images captured at 4 polarization angles (0°, 45°, 90°, and -45°), and the image resolution can reach up to 1280 * 960 pixels. Achieving a reconstruction ratio 1:20. Experimental validation confirms that the spectral reconstruction error consistently remains below 0.14%. Extensive high-dimensional imaging experiments were conducted under indoor and outdoor conditions, showing the system's significant adaptability and robustness in various environments. Compared to traditional imaging devices, such as hyperspectral cameras that could only acquire spectral information, while polarization cameras are limited to polarization imaging, this integrated system successfully overcomes these technological constraints, providing an innovative and efficient solution for high-dimensional optical sensing applications.

## Introduction

Traditional colour cameras can only capture spatial and colour information, but cannot acquire higher-dimensional data such as spectra or polarizations. Current commercial hyperspectral cameras (e.g., IMEC [1]) are capable of spectral imaging. The mainstream approaches are divided into spectral scanning and focal plane array (FPA). Spectral scanning utilizes dispersive devices (e.g., prisms or gratings) to decompose incident light into different wavelengths and sequentially record images for each wavelength. Although this method achieves high spectral resolution, its time resolution is reduced. On the other hand, the FPA approach integrates multiple detectors with narrow bandpass filters on the focal plane, enabling simultaneous recording of multiple images at each wavelength, but has limited spatial resolution. Polarization cameras (e.g., THORLABS [2]) allow polarization imaging by decomposing the incident light into different polarization states, using micro-polarization filters. Each filter corresponds to a specific polarization state, enabling simultaneous recording of multiple linear polarization images with 0°, 45°, 90°, and -45° angles. However, the above cameras fail to simultaneously collect High-dimensional information, such as spatial, spectral, and polarization images.

The rise of computational imaging has opened up new ways to obtain high-dimensional information. In the spectral dimension, Xiong et al. [3] proposed integrating dynamically reconfigurable and reusable metasurface-based micro-spectrometers with CMOS image sensors. Achieving the spatial-to-spectral ratio is 400:400 (450-750 nm) through the recording and reconstruction of modulated incident light intensities. Bian et al. [4] integrated various broadband modulation materials onto image sensors to non-uniformly couple target spectral information into each pixel with high optical throughput, using intelligent reconstruction algorithms, achieving image resolution of 2048×2048 pixels and enabling real-time hyperspectral imaging with the spatial-to-spectral ratio is16:61 (400-1000 nm) and 16:35 (1000-1700 nm). Wu et al. [5] pioneered the combination of a physics-driven deep unfolding network with a generative implicit diffusion model for high-quality hyperspectral reconstruction, achieving a resolution of 660×714 pixels. However, this method was limited to the 450-650 nm range. Zhang et al. [6] developed a novel framework for high-resolution spectral-spatial applications by integrating 16 random spectral filters with a deep neural network (DNN). Achieving a spatial-to-spectral ratio of 16:301 (400-700 nm) with an image resolution of 1280×480 pixels. Wang et al. [7] proposed a scalable random spectral filter based on photonic crystals for spectral reconstruction, the spatial-to-spectral ratio is 9:50 within the 550-750 nm range. Most recently, Motoki et al. [8] used Fabry–Pérot filters placed onto a monochromatic image sensor, achieving a spatial-to-spectral ratio of 64:20 (450-650 nm) with an image resolution of 1920×1080 pixels. Cui et al. [9] proposed a spectral convolutional neural network (SCNN) with matter meta-imaging, by integrating very large-scale and pixel-aligned spectral filters on a CMOS image sensor, the spatial-to-spectral ratio is 9:216 (400-800nm), achieving image resolution of 400*533 pixels, this is the first integrated optical computing utilizing natural light. Despite these spectral breakthroughs, polarization results remain unaddressed. Li et al. [10] proposed a computational hyperspectral imaging (HSI) scheme named SpectraTrack, which achieves high spatiotemporal-spectral resolution with 1200 spectral channels covering 400–1000 nm and an image resolution of 960×960 pixels, but this system needs a scanning imaging spectrometer and an RGB camera result in the system volume is relatively large.

Polarization, spectral fusion technology faces greater challenges. Fan et al. [11] introduced an innovative concept that leverages optical interfaces' spatial and frequency dispersion properties to modulate polarization and spectral responses in the wavelength space. However, this approach is only available with limited targets (e.g., illuminated masks). Zhang et al. [12] developed an encoding metasurface paired with a neural network, enabling a normal camera to acquire hyperspectro-polarimetric images from a single snapshot, which reconstructed 14 polarization states. It achieved a spatial-to-spectral and polarization ratio of 100:1260 (700-1150 nm), but the image resolution was only 42*17 pixels. Wen et al. [13] developed a spectral-polarization camera that combines full Stokes polarization imaging with multispectral imaging, achieving an image resolution of 2016×2016 pixels and enabling polarization reconstruction, However, they placed the polarizers in front of the camera for polarization reconstruction, and the spatial-to-spectral and polarization ratio was limited to 16:28 (400-700 nm). Yet such systems remain constrained by complex system architectures, low computational reconstruction efficiency, and constrained practical applicability limitations. Other spectral imaging technology using narrow bandpass filters has also been continuously advancing, from utilizing MIM oscillators [14], plasmonic nanoresonators [15], silicon nanowires [16,17], and other nanostructures [18].

Building on these foundations, our system achieves unprecedented efficiency in high-dimensional reconstruction, recovering 80 high-dimensional channels (across spectral and polarization domains) from only 4 input channels. This corresponds to a reconstruction ratio of 1:20 in spectral and spatial dimensions, enabling spectral imaging from 400 to 800 nm at 20 nm intervals with an error rate lower than 0.14%. With an image resolution of 1280×960 pixels, our system sets a new benchmark for compact high-dimensional optical sensing systems.

### Design of High-Dimensional Encoder and Decoder（HDE, HDD）

In this study, we propose a high-dimensional spectral polarization imaging system modulated by a self-designed encoder. As shown in Figure 1, the system mainly consists of two parts: High-Dimensional Encoders (HDEs) and a High-Dimensional Decoder (HDD). The HDE effectively encodes high-dimensional cubic matrices in spectral, polarization, and spatial domains, while the HDD design utilizes a High-Dimensional Neural Network (HDNN) based on adversarial networks, which possesses strong feature extraction capabilities, effectively handles complex high-dimensional data, significantly reduces data redundancy, and improves information processing efficiency.

HDEs were designed based on the multilayers. We desire these filters to have a spectrum response under specific polarization states, while they should have nearly 0 transmittance under polarized light with orthogonal states. For example, HDE1 has the transmission spectrum under 0° polarization, as shown in Figure 2a, while it has no transmission under 90° polarization. Eventually, 4 HDEs were designed and their transmission spectra at 0°, 45°, 90°, and -45° were presented in Figure 2, showing their High-dimensional modulation capabilities. These four encoders were selected by calculating the lowest correlation coefficient combinations in the heat map in Figure 3. The correlation coefficient between HDE1 and HDE2 is only 0.00416 (the lowest in the entire matrix), while HDE1-HDE3 (0.06962) and HDE3-HDE4 (0.07483) also exhibit extremely low correlations. Although HDE2-HDE3 (0.33007) and HDE2-HDE4 (0.41521) are slightly higher, they remain significantly lower than correlations involving other encoders (e.g., HDE5-HDE6: 0.60858). This combination optimizes overall low correlation and balance while avoiding high-correlation interference from pairs like HDE4-HDE5 (0.71392) or HDE1-HDE6 (0.649), which would compromise independence. This design ensures complementarity and independence among the HDE in spectral modulation, achieving synergistic optimization of spectral and polarization characteristics and significantly enhancing the overall system performance.

To efficiently decode the acquired 4-channel spectrally polarized high-dimensional images, we propose a High-dimensional neural network based on adversarial networks, combined with the powerful feature extraction and pattern recognition capabilities of deep learning architectures. As illustrated in Figure 4a, it uses a U-Net-shaped architecture as the baseline, the encoder section consists of four sequential down-sampling stages, each consisting of a DSC

Module (Conv3x3 + BN + LeakyReLU) as shown in Figure 4b and a maximum pooling layer. The DSC Module uses a 3×3 convolutional kernel to perform a deep separable convolution, followed by batch normalization (BatchNorm) and a LeakyReLU activation function with a negative slope of 0.2, which reduces the number of parameters and enhances gradient propagation. The convolution operation of each stage multiplies the number of input channels by 4→64→128→256→512 sequences, and the maximum pooling layer compresses the size of the feature map to H/2×W/2, H/4×W/4, H/8×W/8, H/8×W/8 through a 2×2 window and step 2. The decoder part is set up with three upsampling stages, and each stage uses a 3×3 transposition convolutional kernel (ConvTranspose2d) with step 2 to double the size of the feature map (such as H/8×W/8→H/4×W/4) and gradually decrease the number of channels by 512→256→128→80. The hopping connection mechanism fuses the upsampling features of the decoder with the co-scale features of the encoder through channel splicing (Concat), to combine the low-level details and high-level semantic information. Finally, the network maps 4-channel features to 80-channel output. During the decoding process, the interpolation of the data was performed, the interpolation operation enhanced the resolution and fineness of the data, leading to clearer and more accurate reconstructed spectral images, and 20 spectral images containing 4 polarization information were successfully restored, the final output comprises 80 images, with each image corresponding to a specific polarization state and spectral channel, thereby enabling high-precision reconstruction of spectral and polarization information. The multi-channel fusion mechanism ensures the efficient integration of information by leveraging the correlation features between the 4 images. Furthermore, the network optimizes the performance of the generator and discriminator through adversarial training, which significantly improves the reconstruction accuracy and robustness.

Our training dataset was collected from a self-designed optical path shown in Figure 5. A non-polarizing 50:50 beam-splitting cube (LBTEK, BS1455-A) divides the light source into two paths. One path passes through our high-dimensional imaging system. The other path employs a narrow bandpass filter and a linear polarizer (LP), then directly enters a CMOS camera (ONSEMI-MT9J003). Although the beam splitter may introduce some polarization effects during splitting, this issue has been fully considered and corrected during system calibration. Furthermore, to ensure optical consistency, the CMOS camera model and lens parameters (8mm, F2.8) marked in our image are the same as those in our high-dimensional imaging system. The dataset forms 20 × 4=80 feature channels through the full combination of 20 spectral channels (at 20nm intervals within the 400-800nm range) with 4 polarization channels. The model was trained using the Adam optimizer (β1 = 0.5, β2 = 0.999) for 200 iterations. The learning rate was initialized to 0.002, decayed every 20 epochs, and the batch size we set to 32. We chose adversarial Loss and MAE as the loss function. We trained the model on the PyTorch platform with a single NVIDIA RTX 4090 GPU.

## Experiment Result and Discussion

We conducted a series of experiments to validate the quantitative and qualitative performance of our system. First, we examined the spectral reconstruction results and spectral reconstruction errors of the system. We not only compared the synthesized RGB image and the original image, but also compared the reconstructed spectrum (RS) curve with the ground truth (GT) curve. Second, to evaluate the universality of the polarization reconstruction performance, we reconstructed the polarization channels in the indoor environment and the outdoor environment.

Initially, spectra and spectral images of the system were examined. In the initial phase of the experimental program, the ability of the system to reconstruct the spatial is determined by utilizing the visual comparison between the synthesized RGB image and the original image, as shown in Figure 6. Then, the reconstructed spectral data generated by the system is compared and analyzed using independent reference spectral data. 9 points are randomly selected, as shown in Figure 7. By comparing the RS curve with the GT curve, as shown in Figure 8, we can see that there is significant overlap on the curve, and we find that the spectral response of each band is highly consistent with the reference spectrum, which fully validates the excellent spectrum reconstruction capability of the system. In addition, a numerical analysis was performed to quantify the differences between the reconstructed spectral results and the ground truth data. We calculated the maximum spectral reconstruction error and the minimum spectral reconstruction error (RE = (RS-GT)/100%), and the final calculated spectral reconstruction error ranged from 0% to 0.14% ($RE_{max}$=0.14%, $RE_{min}$=0). $RE_{max}$ and $RE_{min}$ are in Figure 8b and Figure 8d, respectively. The error calculation confirms the reliability of the system. This result confirms that the system not only achieves high spatial structure reproduction but also exhibits excellent accuracy and stability in spectral reconstruction, thus highlighting its reliability. The excellent comparison between the synthesized RGB image and the original RGB image, along with the low error between the reconstructed spectrum and the reference spectrum, demonstrates that the system not only excels in spatial structure reproduction but also achieves remarkable accuracy and stability in spectral reconstruction, providing reliable technical support for advanced spectral imaging applications.

To verify the system's ability to recover the polarization state, images were captured from 4 polarization channels (0°, 45°, 90°, and -45°), and the corresponding polarization images were then reconstructed, as illustrated in Figure 9. By analyzing the reconstructed polarization images, it is evident that the brightness variation of the target object varies significantly across different polarization states. These variations, observed through a Pad screen and a polarizing lens, manifest as notable changes in brightness, contrast, and texture in the 0°, 45°, 90°, and -45° polarization images. These results demonstrate the system's ability to discriminate and accurately reproduce light field characteristics in different polarization states. The outdoor environment further demonstrates the robustness and reliability of the system under complex lighting conditions. Despite the presence of dynamic changes in outdoor lighting, the system effectively extracts and distinguishes the light field characteristics of different polarization states, as shown in Figure 10, We can observe the variations in light intensity and contours of the metal signboards signboard (0° and -45° are bright, but 45° and 90° are dark), car window glass (0° and 45° are bright, but 90° and -45° are dark), and the metal signboards in the upper right corner (0° and -45° are bright, but 45° and 90° are dark). This further proves the system's applicability and stability in real environments. These advantages offer considerable potential for the system in outdoor detection, target recognition, and other practical applications. Table 1 provides a comparative analysis of various spectral/polarimetric imaging techniques.

## Conclusion

The proposed high-dimensional imaging system integrates four low-correlation encoders (HDE 1-4) to capture encoded image data from the target object. This data is then demodulated and reconstructed by a high-dimensional neural network (HDNN) based on adversarial networks. Experimental verification has confirmed that the system is capable of simultaneously acquiring 20 sets of spectral images in the 400-800 nm wavelength range, each containing information from four polarization directions (0°, 45°, 90°, -45°), totaling 80 images and achieving an efficient reconstruction ratio of 1:20. The system achieves an image resolution of 1280x960 pixels and a spectral reconstruction error of less than 0.14%. It has demonstrated excellent cross-scene generalization performance, as evidenced by polarization imaging experiments conducted in diverse indoor and outdoor test environments. The successful construction of this innovative high-dimensional spectral polarization imaging technology system effectively resolves the inherent contradiction between information dimension and imaging efficiency in traditional imaging systems. The results underscore the system's broad application prospects in key technical domains such as biological microscopic imaging, remote sensing with fine detection, and material multi-parameter analysis.

**Funding.** This work was supported in part by the National Natural Science Foundation of China (Grant No. 42201336).

**Disclosures.** The authors declare no conflicts of interest.

**Data availability.** The data underlying the results presented in this paper are not publicly available at this time but may be obtained from the authors upon reasonable request.

**Author contributions.** Y. Y. and Z. G. contributed equally to this work. X. L. and X. H. conceived the study and designed the overall framework. Y. Y. and Z. G. developed the HDNN framework and conducted the experiments. X. H. supervised the project. P. B., W. S., and R. Y. provided optimization suggestions. R. K., X. X., C. Q., and J. M. contributed to the figure design. Y. Y., Z. G., P. B., and X. H. wrote the manuscript with contributions from all other co-authors. All authors reviewed and approved the final manuscript.

## Figures

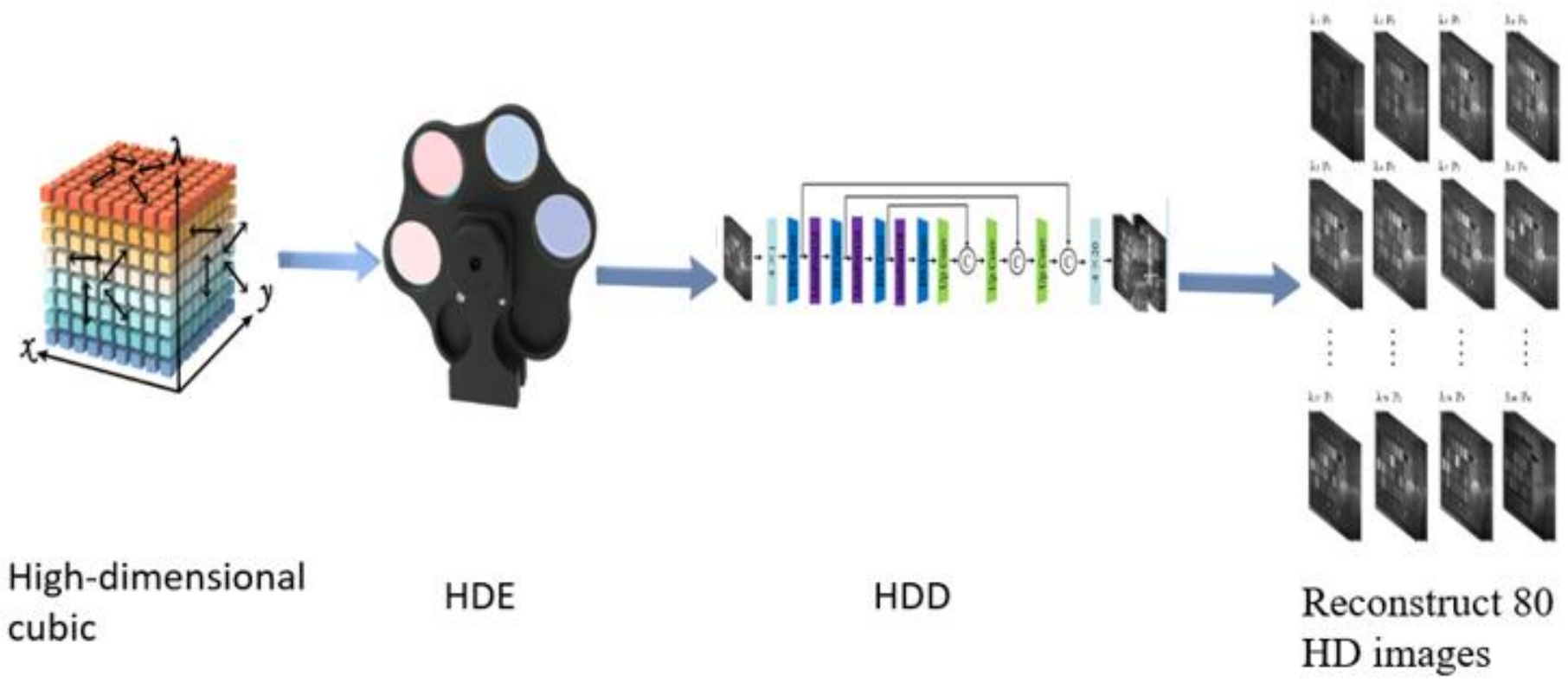

Fig.1 The schematic diagram of the high-dimensional encoding and decoding process, the high-dimensional cubic is encoded by HDE and decoded by HDD to reconstruct 20 spectral datasets. Each dataset contains images captured at 4 polarization angles (0°, 45°, 90°, and -45°).

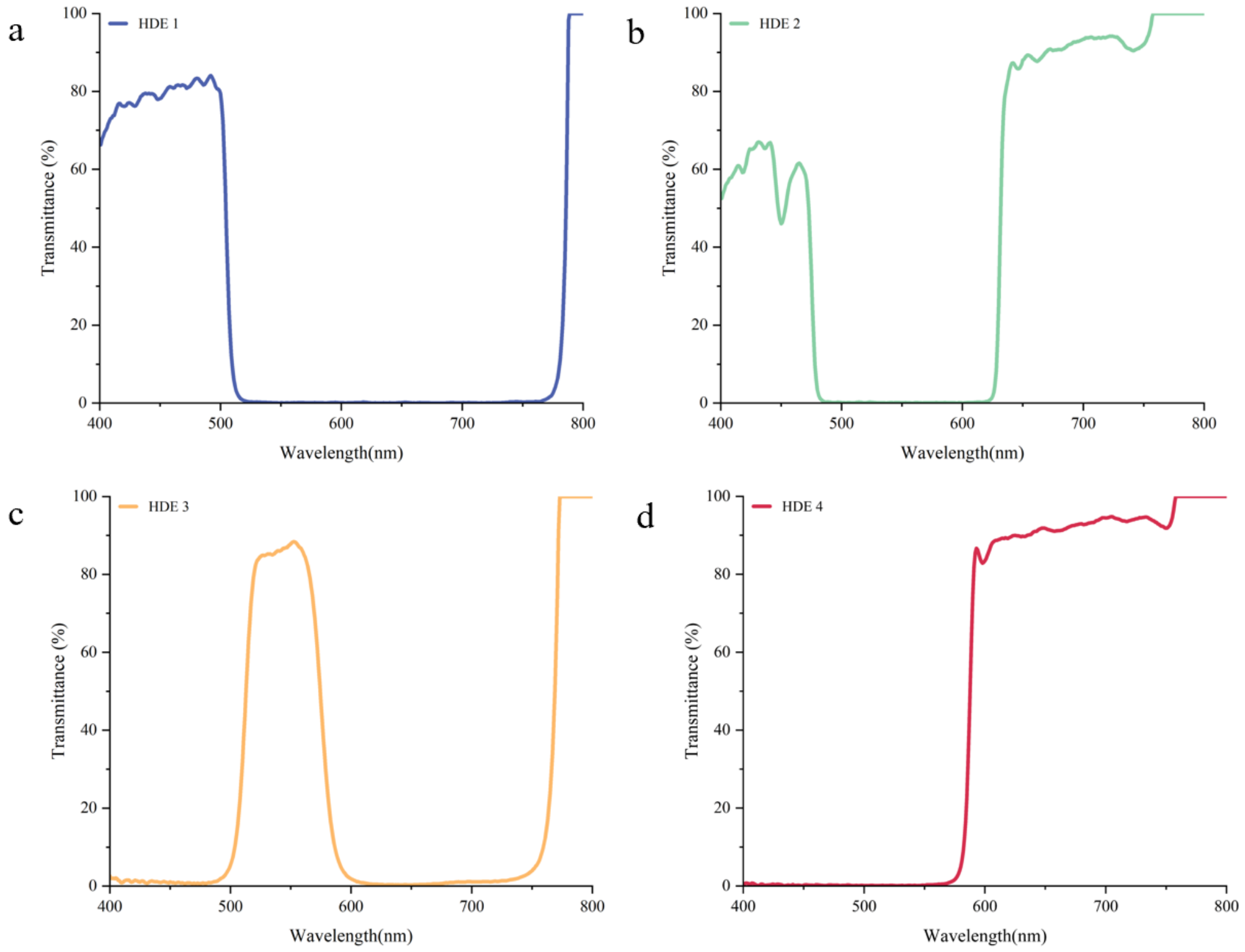

Fig.2 4 selected high-dimensional encoders, HDE1, HDE2, HDE3 and HDE4 modulate 0 °, 45 °, 90 °, -45 ° polarization and spectra information, respectively.

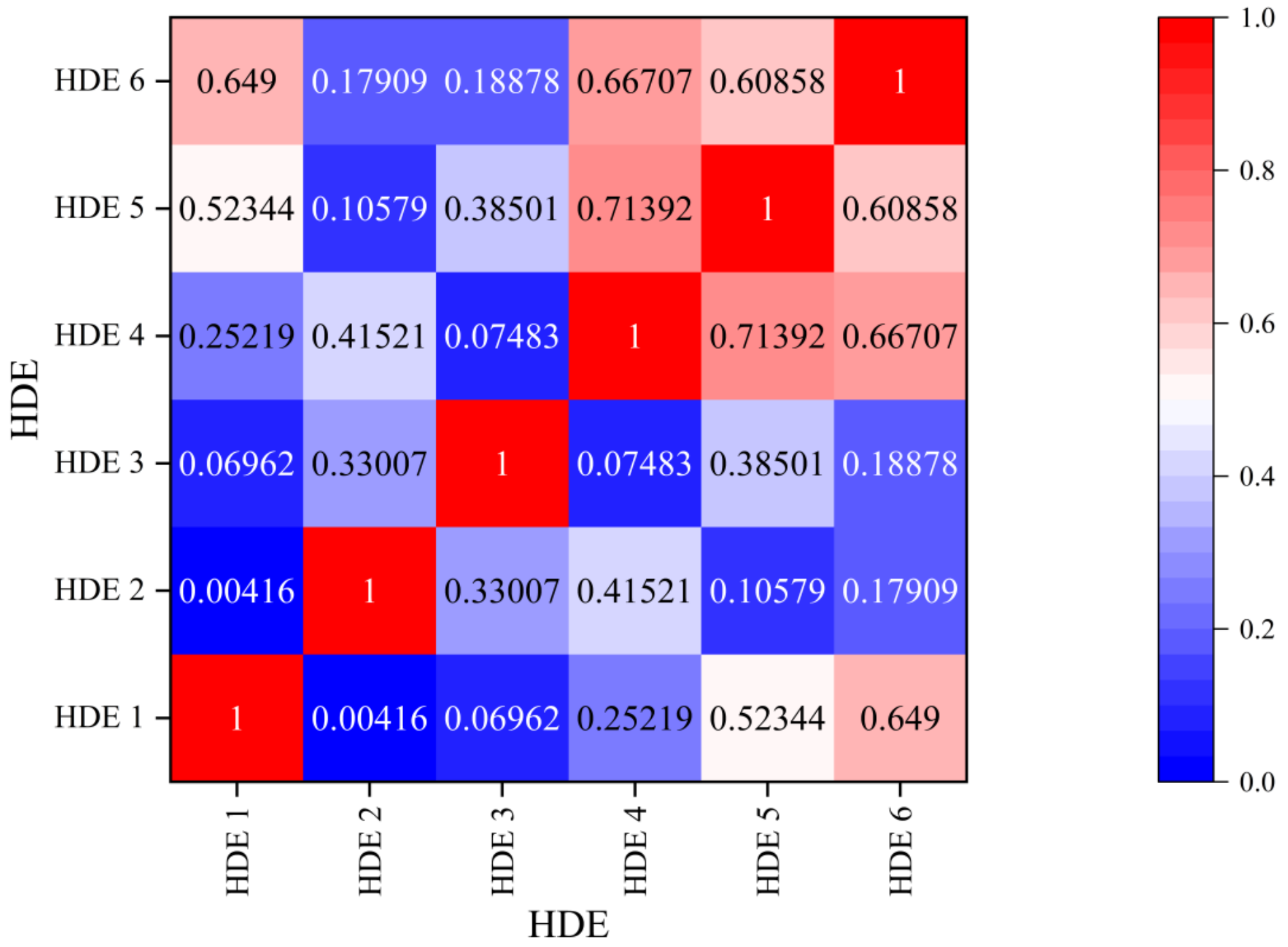

Fig.3 Thermal map of 6 encoders, based on the heatmap data, the combination of HDE 1, HDE 2, HDE 3, and HDE 4 is selected as the four least correlated encoders. The correlation coefficients between HDE1-2 (0.00416), HDE1-3 (0.06962), and HDE3-4 (0.07483) show extremely low correlations. Slightly higher yet still significantly weaker correlations are observed in HDE2-3 (0.33007) and HDE2-4 (0.41521) compared to other encoder pairs (e.g., HDE5-6: 0.60858). This configuration maintains overall low correlation and balance while effectively avoiding high-correlation interference items (such as HDE4-5: 0.71392 and HDE1-6: 0.649).

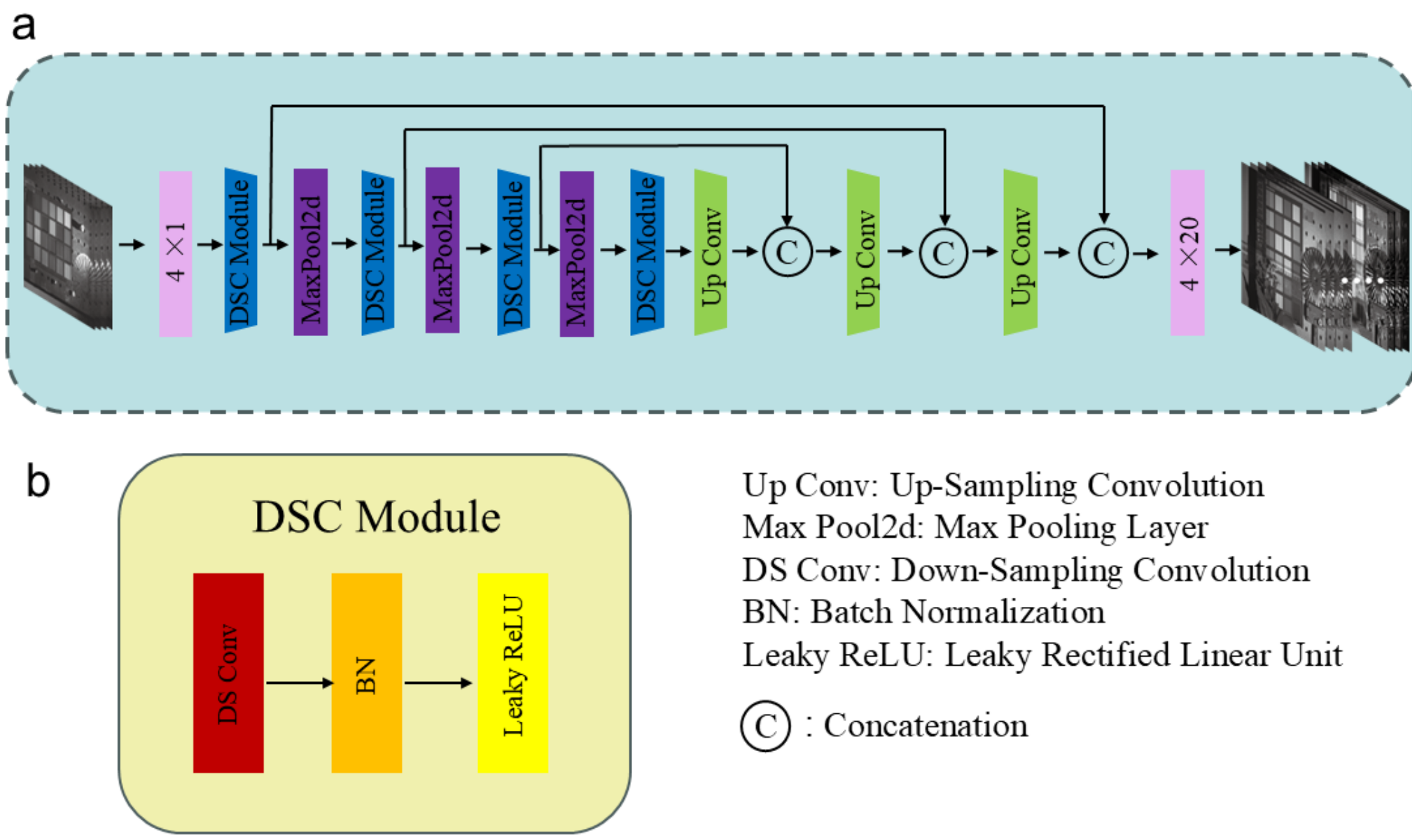

Fig.4 The framework of a high-dimensional neural network model

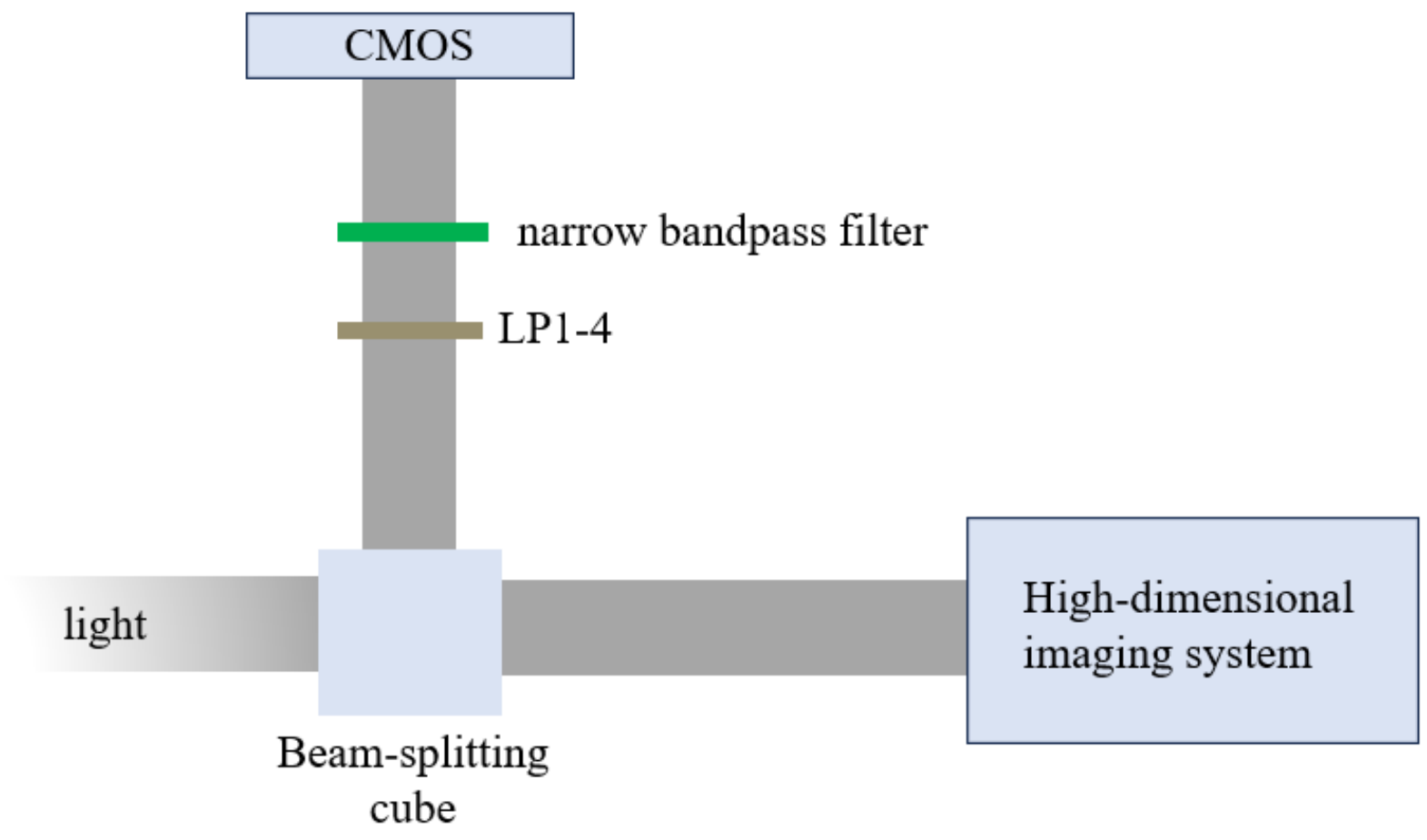

Fig.5 Schematic of polarization spectrum image acquisition, the Beam-splitting cube splits the incoming light into the data acquisition system(including LP: Linear polarizers, narrow bandpass filter, and CMOS camera) and our high-dimensional imaging system.

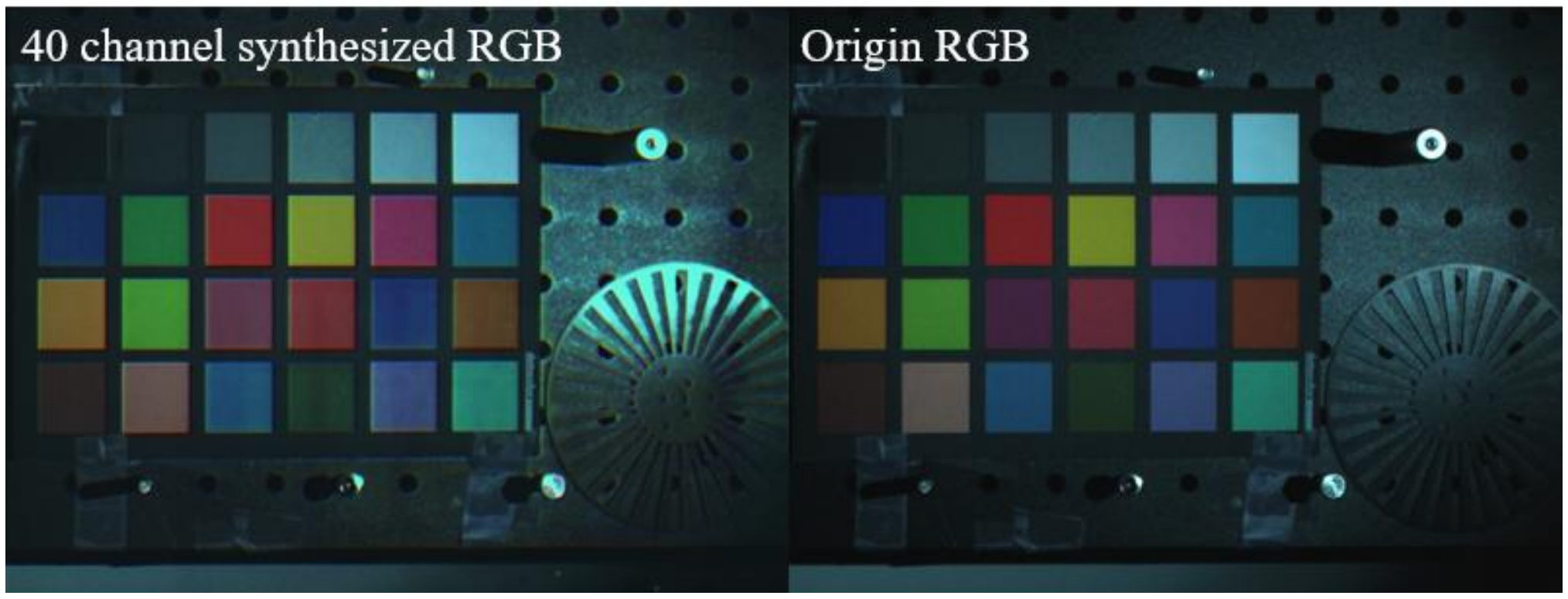

Fig. 6 Original RGB image(left) and 40-channel synthesized RGB image(right)

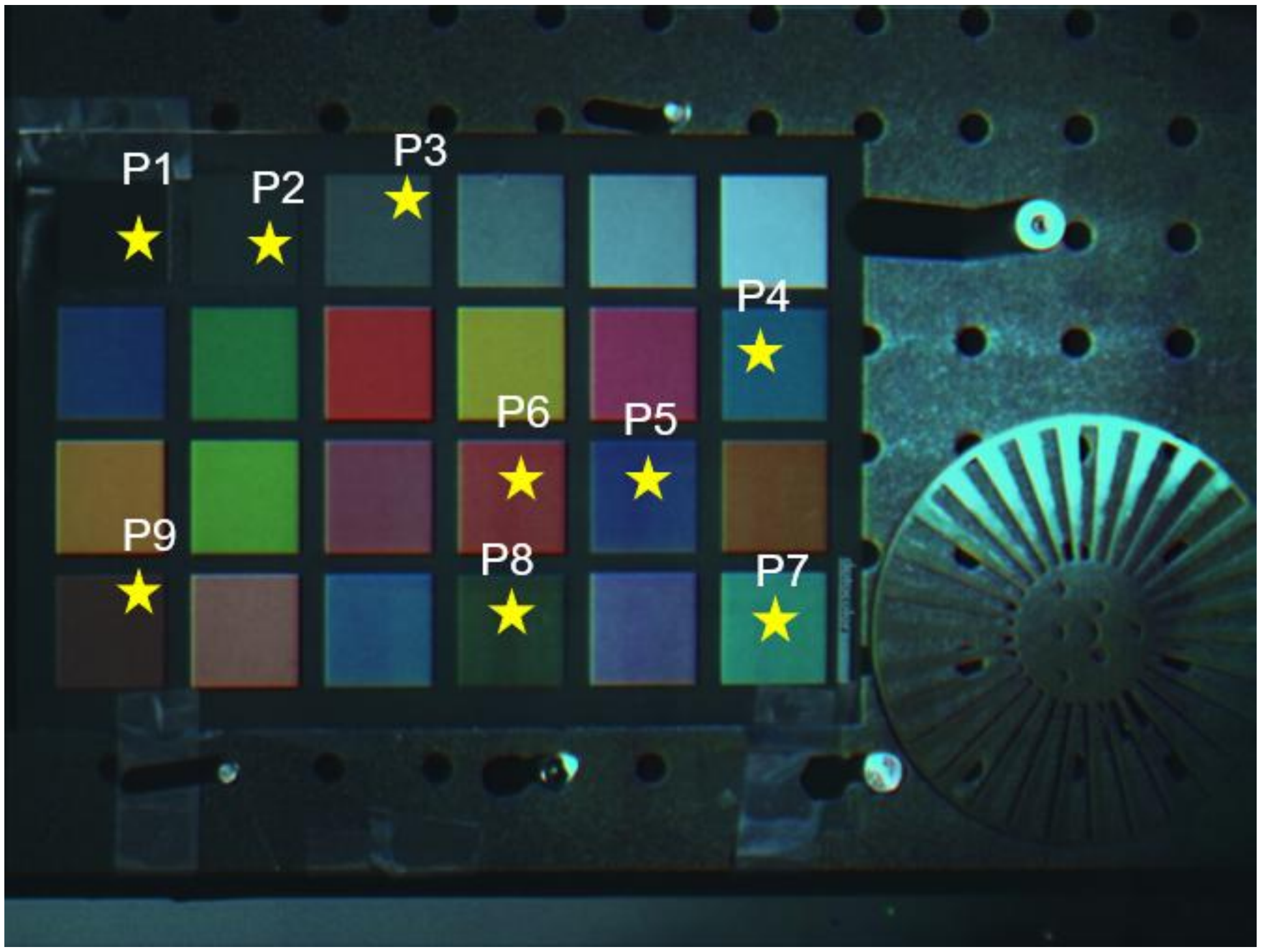

Fig.7 9 randomly selected hyperspectral data points.

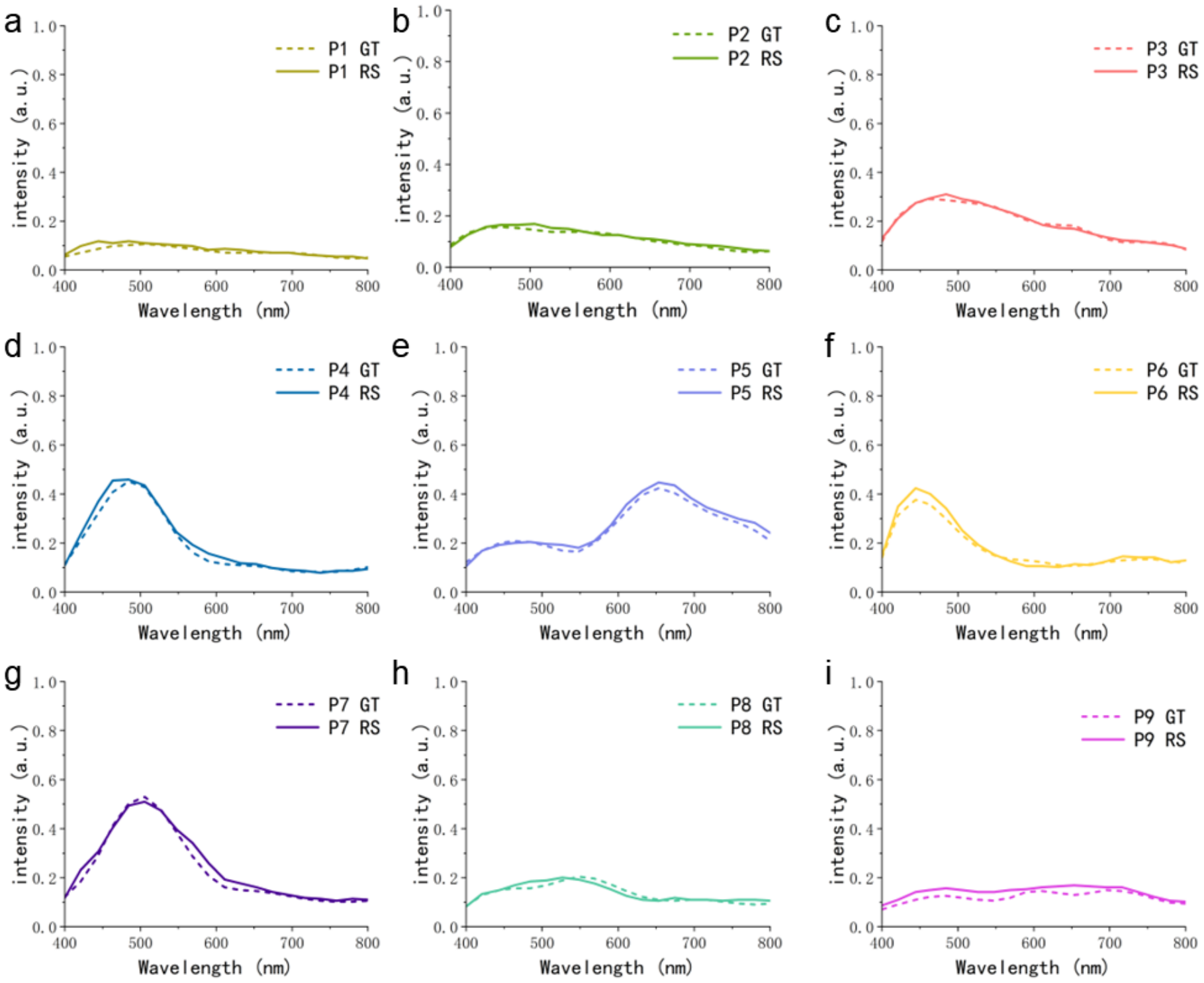

Fig.8 The spectral comparison between the reconstructed spectra (RS) and ground truth (GT), Points P1 to P9 in Figure 8 correspond to subfigures (a) to (i) in the figure. $RE_{max}$ and $RE_{min}$ are in b and d, respectively

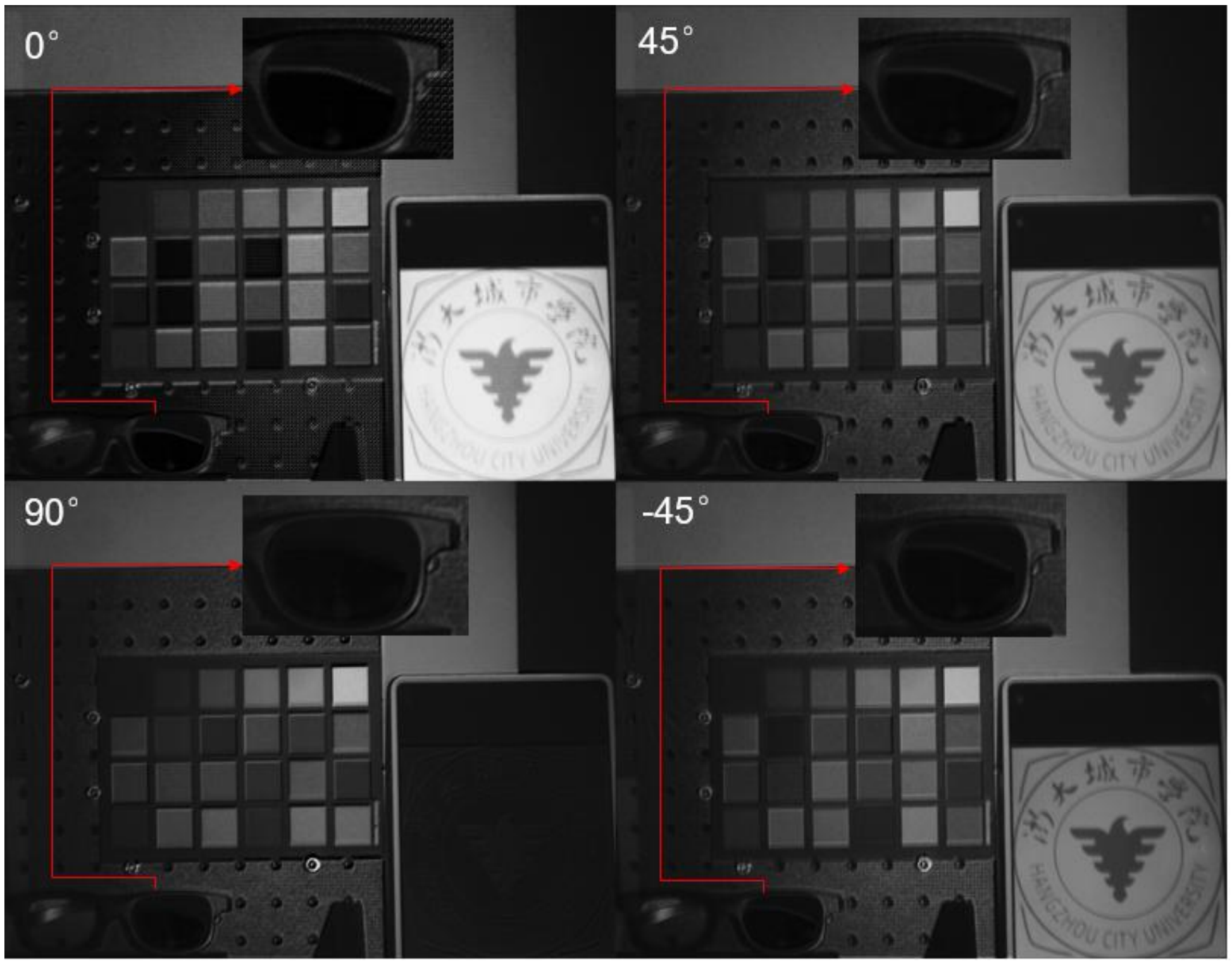

Fig.9 Reconstructed 0°, 45°, 90°, and -45° polarized images (indoor). The sunglasses frame and Pad screen are noticeable at 0° and 45° polarization, but the sunglasses frame is barely visible at 90° and -45° polarization, and the pad screen is noticeably darkened.

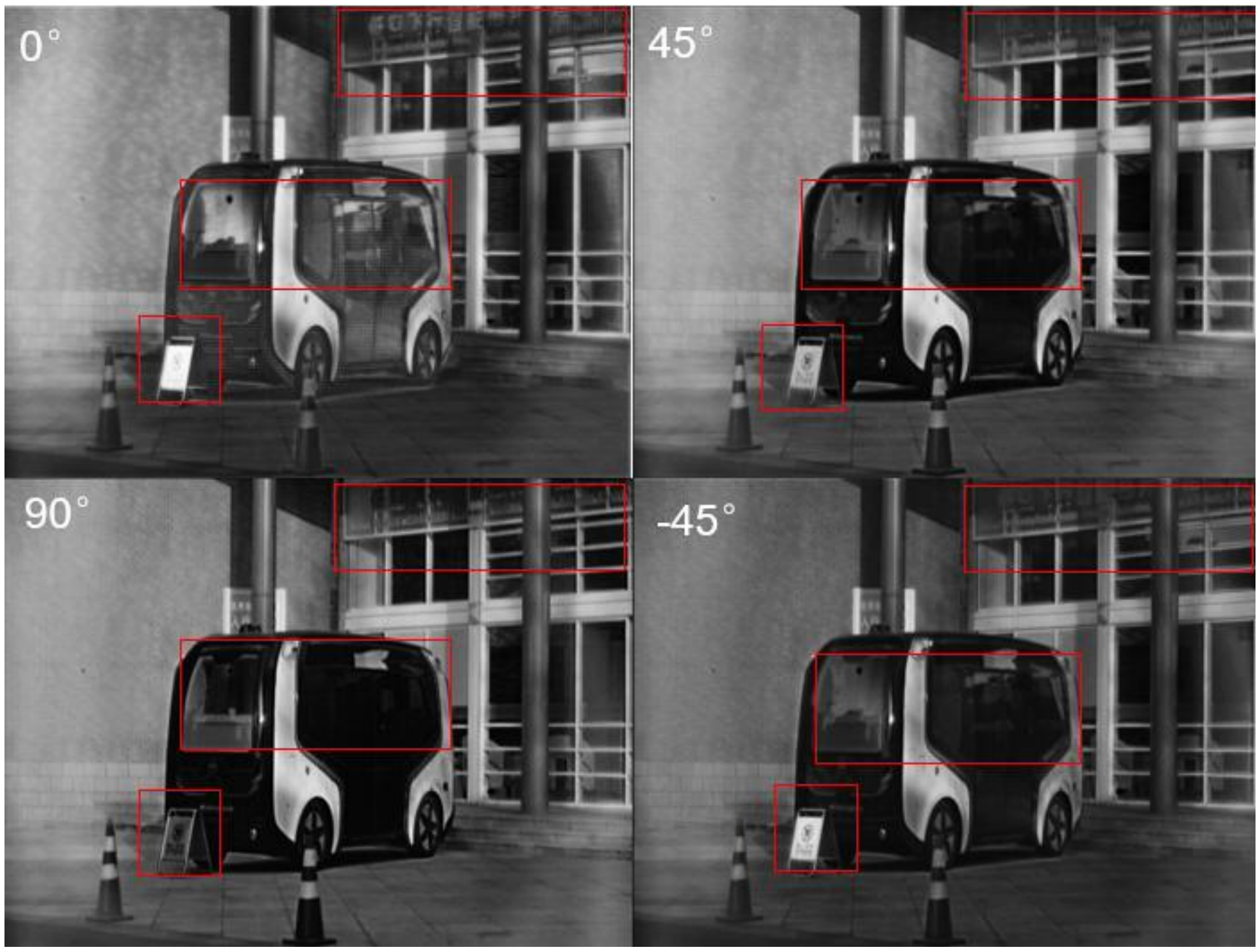

Fig.10 Reconstructed 0°, 45°, 90°, and -45° polarized images(outdoor). We can observe the variations in light intensity and contours of the metal signboards signboard (0° and -45° are bright, but 45° and 90° are dark), car window glass (0° and 45° are bright, but 90° and -45° are dark), and the metal signboards in the upper right corner (0° and -45° are bright, but 45° and 90° are dark).

| **Works** | **Spatial to spectral/polarization ratio** | **Reconstructed Optical Dimensions** | **Operating wavelength (nm)** | **Imaging or measurement** | **Image Resolution** |
|---|---|---|---|---|---|
| Our work | 1:20 | 4 | 400-800 | Imaging and measurement | 1280*960 |
| Optica Xiong et al. (2022)[3] | 400:600 | 0 | 450-750 | Measurement | - |
| Nature Bian et al. (2024)[4] | 16:61-16:35 at different wavelengths | 0 | 400-1700 | Imaging and measurement | 2048*2048 |
| Nature Fan et al. (2024)[11] | - | 5 | 400-900 | Imaging and measurement | None provided |
| Nature Wu et al. (2024)[5] | - | 0 | 450-650 | Imaging and measurement | 660*714 |
| Light Zhang et al. (2021)[6] | 16:301 | 0 | 400-700 | Imaging and measurement | 1280*480 |
| Nature Wang et al. (2019)[7] | 36:200 | 0 | 550-750 | Measurement | \ |
| Laser Wen et al. (2025)[13] | 16:28 | 6 | 400-700 | Imaging and measurement | 2016*2016 |
| Nature Motoki et al. (2023)[8] | 64:20 | 0 | 450-650 | Imaging and measurement | 1920*1080 |
| Science Advances Zhang et al. (2024)[12] | 100:1260 | 14 | 700-1150 | Imaging and measurement | 42*17 |
| Nature Li et al. (2024)[10] | - | 0 | 400-1000 | Imaging and measurement | 960*960 |
| Nature Cui et al. (2025)[9] | 9:216 | 0 | 400-800 | Imaging | 400＊533 |

Table 1: Comparison of various spectral-polarization imaging technologies